# Usability as a Dominant Quality Attribute

A. Raza and L. F. Capretz
Department of Electrical and Computer Engineering
University of Western Ontario
London, Ontario, Canada N6A 5B9

*Abstract -* *Whenever an architect or a team of architects begins an architectural design, there are certain goals set to achieve. There are many factors involved in setting up goals for the architecture design such as type of the project, end user perspective, functional and non-functional requirements and so on. This paper reviews and further elaborates strategy for the usability characteristics of software architecture. Although user centered designs are tremendously gaining popularity, still in many design scenarios, usability is barely even considered as one of the primary goals. This work provides an opportunity to compare different strategies and evaluate their pros and cons.*

## 1 Introduction

Software architecture is not only influenced by technical, business and social factors rather architecture is in fact the result of these factors [1]. Being the vehicle of all the stakeholders' communication, it provides a common language platform to express their demands, concerns and negotiations. Throughout architectural design, implementation and deployment, different quality attributes like availability, variability, performance, maintenance, testability, modifiability, scalability and usability are considered, with some having more preference on the others, depending upon the nature of the application for which the architecture is being designed. This also provides the basis of trade-offs a software architect has to make during the life cycle of an architecture.

If we look carefully and cautiously, we will find one common factor, always there. This common factor is "end user." Whether we are considering the business and the social factors or we are taking care of stakeholders' concerns or we are deploying the architecture to the client or we are marketing our final product based on this architecture, in each and every case we are dealing with people. This leads to a question that why should not our design be "people oriented" instead of "technology oriented"? And why not usability should be the dominant quality attribute?

This work investigates the following research question:

RQ-1: Is it pragmatic at all to consider usability as one of the primary quality attribute in almost all architectural designs?

Firstly, here are two examples of bad designs [2]:

Example – 1: This student ID card has three numbers. However, none of the numbers are labeled.

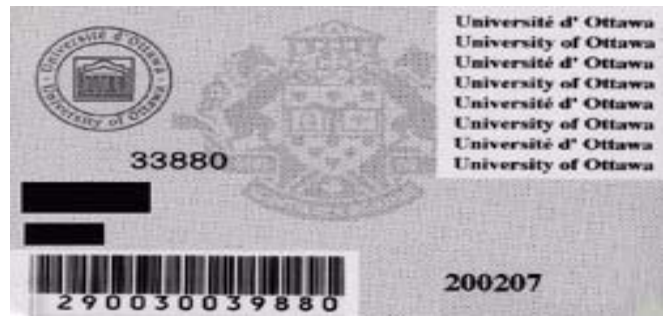

Design suggestion: The student ID number, being the most frequently accessed number on the card should be labeled and it should be the most prominent.

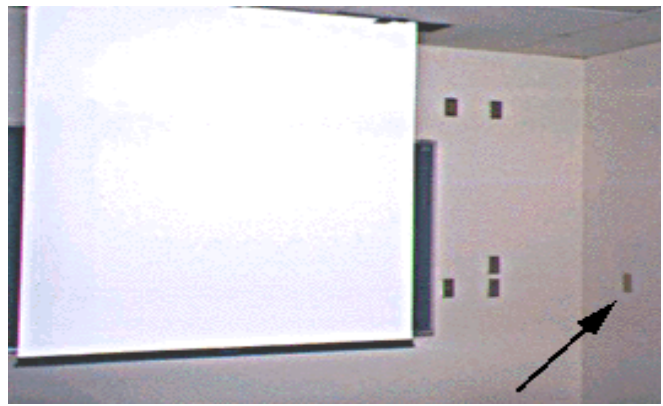

Example – 2: This picture shows a projector screen lowered from the ceiling in front of a blackboard. To the right of the screen are some electrical switches. Guess which switch raises and lowers the screen? Is the switch next to the screen? No, it is the switch farthest away from the screen!

Design suggestion: Place a control next to the device it controls.

Usability Professionals' Association (UPA) has number of real life success stories where usability helped in improving a project [3].

One of those is about U.S. Automobile Association (USAA), which provided ergonomics training to more than 20,000 of its employees and did usability testing of all software they developed or purchased for use. Through the use of ergonomics programs, the company not only improved its productivity but significantly reduced training costs and workers' compensation claims.

In another case study, in 1991, the Ford Motor Company's car dealers were facing some trouble with their accounting system. They conducted usability study of the system, on the basis of the which, specific changes were made. Not only that this resulted in significant savings for the company but the calls to the help line were dropped to almost none as well.

## 2 Literature Review

Te'eni [4] highlights that for thirty years, organizational tasks and context have been at the centre stage in management information systems, it is thus the time to research what the higher level tasks, users wish to do and how users interact with the computer to achieve these tasks. Zadrozny et al. [5] stress upon natural language dialogue for personalized interaction. Personalization into the dialogue system means dealing with many factors. However, considering user's profiles and preferences, the requests can even be understood in a better way and by avoiding misunderstandings. Natural language personalization can also mean customizing to a style or to a group of users, not necessarily an individual. We need to design repositories of textual knowledge and some natural language relational databases. The knowledge has to be made adaptable to the user. Although it involves changing the format in which the data is stored as well as creating dialogue interfaces. Still current technologies may just be extended to solve the issue [5].

Lindgaard [6], however strongly argues that although thoroughness, efficiency, and validity are necessary to establish the effectiveness of a testing procedure, they are irrelevant in human computer interaction practice. Marc Chrusch [7] highlights some myths such as usability only increases development costs and lengthens the development time, the user interface is really just adding good graphics to make the application appealing, usability is user interface design, usability is really just common sense, as long as developers are familiar with guidelines good user interfaces will be designed, usability testing is not needed if the development team has been working with the users a long time and knows what they want, and we will handle that in the help / training / documentation. The author explains seven myths about usability and how they can be encountered.

Seffah and Metzker [8] also highlight the obstacles and myths of usability and software engineering (SE). The user centered designs techniques, developed by Human Computer Interaction (HCI) teams are generally found difficult to understand by software development organizations, as they have techniques of their own, even the usability techniques. There could be a harmonious coexistence between the two communities if the obstacles in their way could be pointed out and be avoided [8]. A forum is thus needed to share the ideas between the two fields such that a cohesive relationship can be built between HCI and SE practices and research.

Koppelman and Dijk [9] focus on the role of clients and users in projects, how to learn to deal with different stakeholders who look at the product in a different perspective, how to communicate with them and how to involve the real users and clients in the design process. They suggest that the designers should not simply rely on their own experiences and instincts.

Henderson [10] is curious that why is it so that a company does research, for example research on user centered innovation, but it is found difficult to implement such ideas on the products ? He then comes up with the reason that just concentrating on user needs is not good enough.

Czerwinski and Larson [11] in their research, look at the novel trends and the research techniques that might be useful for the future professionals, especially concerning web designs. A web site can be more acceptable if it provides information to the users of different computer usability background and experience. All new web sites arrange a large set of information either automatically or in a pattern to be easily managed by a user. According to them, one thing is thus for sure that usability will be the tool towards acceptance of any future web design.

Shneiderman [12] looks at the problem of how can computer resources be made available and usable for every one. Designing for any expert computer user is difficult as it is, let alone to design, such that any one can use it. He states that lowering cost of hardware and computer accessories is giving access to more people, but still interface and information design has to play its role. His paper focuses on three challenges in universal usability within web design: technology variety (about hardware, software and network support), user diversity (concerning users of different age, gender, background and disabilities) and bridging the gap (what they already know and what they need to know).

In their research paper, Myers, Weitzman, Andrew and Chau [13] point out those modern applications have many features and dependencies that are helpful most of the times, but sometimes they can be inexplicable even to the expert users. To answer such queries, an application framework called Crystal (Clarifications Regarding Your Software using a Toolkit, Architecture and Language) has been created. The "Crystal" builds question menus dynamically, by current state of the application. It provides invisible objects under every point in the window so users can ask questions by pointing anywhere even the blank space.

Faulkner and Culwin [14] suggest more interaction between human computer interaction and computer science disciplines by adopting HCI as the underlying principle to the systems development.

According to Rosson, Carroll and Rodi [15], the main challenge in teaching usability engineering is to come up with realistic projects for the students, such that meaningful issues could be addressed in a manageable time of a semester.

Zhang [16], however, initiates the discussion of exploring opportunities in college education to make future managers and executives, more human-centered, usability-sensitive and HCI competent.

## 3  Are All The Architectural Designs Made For The People?

This is a big question in the context of our discussion. Instead of finding its answer straight away, we will look at the picture from the other side. We try to find any architecture whose immediate audiences are not human beings. This leads to another question: Can we find any such common architecture?

We can find few odd cases of systems, designed such that they are not directly exposed to the end users, like the ones which may take input from one machine and provide the output to the other; or the architecture of any component or a layer of an operating system, which might not be addressing the end user directly.

But we cannot simply eliminate the human factor even out of these scenarios. Firstly are these architectures not going to affect human beings, ultimately? Secondly we shall not forget the factor of different stakeholders, which are human beings, and they are involved, affect and get affected by the architectural design in one way or the other.

However, even if we buy this idea for the time being, that not all the systems are going to be delivered to the end user and hence will not be used by them, directly; so why waste resources in terms of time, money, and efforts to make them "usable"? Still what will be the percentage of the systems targeted to be designed for the user and will be used by them directly, as compared to the cases mentioned above? Answer is simple. Most of the systems are designed for the users of different domains. And they can be considered successful only if they come up to the expectations of their audiences. Outside the so called "computer community", we find the people either not very comfortable with the computer software or using them in a very limited way.

## 4  Obstacles in the Way of Human-Centred Designs

We have seen rapid changes rather revolutionary advancement in the computer industry, both at the hardware as well as at the software fronts, in the past few decades. Computer hardware and accessories are much cheaper and accessible as they could ever be imagined earlier. No domain of life we can find that has been left uninfluenced by the software. For software developers and programmers, all sorts of visual languages and tools are available to come up with a software solution, practically in the shortest span of time. Then what are the causes of complaints regarding software designs particularly from usability point of view. It appears as if somewhere priorities have not been set right. A very common problem with the programmers and software developers is that while designing they think that if a design is good enough for them to be used, so will it be for the targeted user. In other words, they consider themselves as the end user.

Shneiderman [12] highlights three challenges in attaining universal usability, namely technology variety, user diversity and gaps in our knowledge. Seffah and Metzker [8] argue that "usability" is a confusing concept and its consistent definition has not been agreed upon either by the researchers or by the software developers or even by the standardization organizations. So obviously when a term has not been defined consistently and its definition is not agreed upon unanimously, how could its related issues be addressed appropriately?

Henderson's [10] point of view is slightly different. According to him, stand alone research and suggestions of human computer interaction has not and cannot contribute enough and play its role effectively in the software development.

## 5  Our Suggestions

Keeping in view the above discussion, here are few suggestions which might be helpful in making software designs more user friendly, particularly in the long run.

### 5.1  College Education

Instead of stressing the software developers to be more user-centered, attention needs to be paid towards the computer science and software engineering curriculum and teach the students the importance of usability, and ways and techniques to achieve it. It would be more useful as a part of the long term solution. Courses, particularly the programming languages courses keep telling the principle of "divide and conquer", without making the students realize that whether or not it has any effect on their design. Students shall be encouraged to highlight the aspect of usability, in whatever design they make. They should be kept reminded that just coming up with a programming solution is not good enough; they have to design it for a real user. This will help in nurturing in them a sense of real life designing.

## 5.2 Usability at the Software Architecture Design Level

The approach that architecture has to be more focused on the so called "actually important" software quality attributes, and the usability can always be incorporated at the design level, is one of the root causes of software lacking its credibility in this domain. Architecture is the foundation on which whole building of the software development is based. If only priorities are set right at this stage, by giving usability its due share of importance, we can come up with much better user centered designs.

## 5.3 Programmers Are not the End Users

The earlier this fact is realized by the programmers and software developers that actual users of the software will be different from the programmers, the better it would be. It is a very general belief that a programmer would think of him or herself as a user while designing or developing a software system. If only he/she is actually convinced that this software is going to be used by some other human being, which can be anyone, but them, it could result in incorporating many more aspects and minor details into the design. By realizing this fact that a programmer or designer is not the actual user, brings a different view angle in front of us to look at the things, which can help us in thinking and designing from a perspective, one that of the actual user's, not that of the programmer's.

## 5.4 Collaboration between HCI people and Software Architects

Designing architecture of any software system is the stage when incorporating changes is much easier. If we compare any change suggested by an HCI expert at a stage when all decisions are being made and nothing is finalized yet, to the stage, when architecture has been finalized, stakeholders have been communicated, system design has been implemented and is about to be deployed, the problem picture becomes clearer. So, the sooner we take usability considerations, with the consultation of HCI professionals into account, the better it would be for the project under consideration.

## 5.5 Influences of Other Disciplines

The fact that computer science and software engineering domains do not exist in isolation as per their dealings of software architecture and system development, is very important to realize. This is even more applicable in today's world of internet connectivity and close communication. So the influence on software development particularly from the disciplines like psychology, philosophy, sociology etc, which deal with the human beings, their feelings, their intended actions and reactions and so on, must be taken into account while designing and developing any software. The knowledge and application of these disciplines may give a new orientation to a project altogether, which might not be possible to conceive otherwise, for a computer scientist.

## 5.6 Special Users

A standard user who is young, energetic, knows everything (or at least the basics) to execute any new software may not be our subject user all the times. Users like elderly or children or people having some sort of disability, should also be given serious and enough consideration while designing a system. And how serious and enough the consideration has been, it must be reflected in the design.

## 5.7 Designing Help Features

If "help" features are designed in their literal meanings to provide help to all sorts of users, not only to the expert ones, then it can "help" solving the usability problem. But again to provide help, we need to know exactly what sort of help is actually expected and may be asked for. And this is only possible if the architect and the designer put themselves in the shoes of the actual user. The help shall be made as interactive as possible.

## 5.8 Incremental Approach

Gaming software are good examples of a design with an incremental approach. There are levels of difficulty in almost every computer game. User completes one level and moves on to another with more challenges to face on. The same technique may be applied to any software. Every feature or trick needs not be disclosed to every user, as it may not be his/her requirement, especially for a basic user. User may be encouraged to explore advance features at an advance level.

## 6  Discussion

As Te'eni [4] mentions that due to more rapid advancement in information technology (IT), more research may have been technology based; there is a need of a better theoretical research in the areas like decision support and user satisfaction. However, disagreeing with Lindgaard [6], we believe that thoroughness, efficiency, and validity are necessary to establish the effectiveness of a testing procedure and they are very much relevant in any software practice, which involves people. We fully support Henderson's point of view [10] that an idea can become reality if it involves other stakeholders who participate in the conversion of the concept into a product, too. Just as Shneiderman [12] suggests that encouraging the use of computers to every one including the users with some disabilities, could end up being beneficial to everyone.

Keeping such points of views in mind, our suggestions of stressing user-centered designs at college level education, considering "usability" as a primary quality attribute to achieve at the architectural design level, involving all the stakeholders right from the initial design phase, considering special users' needs, and provision of specific and interactive help could lead towards a path of more user friendly software development.

The crux of the discussion is that usability is one of those quality attributes of a software architecture design, which is always there to be considered, no matter what type of the system is going to be designed.

Few suggestions have been made in this regard, which by no means are complete. The intention was only to look for some long term solution of the stated problem. The suggestions, however, can only be useful once they are implemented in their true letter and spirit.